# High gain two-stage amplifier with positive capacitive feedback compensation

*Alireza Mesri, Mahmoud Mahdipour Pirbazari, Khayrollah Hadidi, Abdollah Khoei*

*Microelectronics Research Laboratory, Urmia University, Urmia, Iran*

*Email: A.mesri1369@gmail.com*

**Abstract:** A novel topology for a high gain two-stage amplifier is proposed. The proposed circuit is designed in a way that the non-dominant pole is at output of the first stage. A positive capacitive feedback (PCF) around the second stage introduces a left half plane (LHP) zero which cancels the phase shift introduced by the non-dominant pole, considerably. The dominant pole is at the output node which means that increasing the load capacitance has minimal effect on stability. Moreover, a simple and effective method is proposed to enhance slew rate. Simulation shows that slew rate is improved by a factor of 2.44 using the proposed method.

The proposed amplifier is designed in a 0.18um CMOS process. It consumes 0.86mW power from a 1.8V power supply and occupies 3038.5$\mu m^2$ of chip area. The DC gain is 82.7dB and gain bandwidth (GBW) is 88.9 MHz when driving a 5pF capacitive load. Also low frequency CMRR and $PSRR^+$ are 127dB and 83.2dB, respectively. They are 24.8dB and 24.2dB at GBW frequency, which are relatively high and are other important properties of the proposed amplifier. Moreover, Simulations show convenient performance of the circuit in process corners and also presence of mismatch.

## 1 Introduction

Operational amplifiers are one of the main building blocks in analog circuits and are widely used in signal conversion, consumer electronics, and communications. Single-stage amplifiers used to be a desirable choice for designers when old CMOS technologies were used. It was mainly because of their high speed, medium to high gain, and stability even with large loads. Gain and output swing of single-stage amplifiers are reduced in modern CMOS technologies, due to reduction of intrinsic gain of transistors and also power supply voltage. Different methods have been proposed in the literature to improve gain in single-stage amplifiers, such as increasing the transconductance and using positive feedback [1-8]. In these methods, phase margin (PM), output swing or linearity are often traded for gain. However, they still fail to deliver the desired high gain in modern CMOS technologies.

Therefore, to achieve high gain and high swing together, two-stage and three-stage amplifiers started to draw attention. These amplifiers usually have multiple poles and zeros and need frequency compensation. Miller compensation is the most conventional method. The compensation capacitor introduces a right half plane (RHP) zero which degrades PM [9-11]. Voltage buffers and current buffers are placed in series with the compensation capacitor to eliminate effect of the RHP zero. Voltage buffers limit the output swing, which is a major problem in modern CMOS technologies. Current buffers on the other hand, reduce the voltage gain, increase offset and noise and sometimes require higher current consumption [12]. Moreover, some other methods have been proposed in the literature that reduce the required compensation capacitance which are useful for heavy capacitive loads [13-16]. The amplifier proposed in [16] is stable for a specific range of load and closed loop gain. In contrast to Miller compensated amplifiers whose $PSRR^+$ approaches 0dB around the gain bandwidth (GBW), the amplifier in [16] improves $PSRR^+$ by 21.6dB at GBW, in addition to broadening of $PSRR^+$ bandwidth. Having high $PSRR^+$ at high frequency is particularly important in mixed-signal systems (e.g. data converters and switched capacitor (SC) filters) with fast digital sub-systems which generate broadband noise [16, 17].

Also, bulk-driven techniques are widely used to design amplifiers for lower supply voltages [18, 19]. Although low voltage operation is an important advantage of these amplifiers, increased area, larger input capacitance, larger input referred noise, lower gain, and lower cut-off frequency are some of their disadvantages [18]. A three-stage bulk-driven amplifier is proposed in [18] which has a high gain. However, the amplifier is prone to instability if the load capacitance is increased. Moreover, the main disadvantage of this amplifier is its supply voltage limit which is 0.5V. This is to prevent forward biasing the bulk-source junction.

A novel high gain two-stage amplifier has been proposed in this paper. The amplifier is compensated by a positive capacitive feedback (PCF) around the second stage and does not need a resistor or any additional circuit in series with the capacitor to feed back the output signal .The compensation capacitor introduces a left half plane (LHP) zero which cancels effect of the non-dominant pole at output of the first stage, considerably. Consequently, the dominant pole happens at the output node and stability has low sensitivity to load capacitance. Moreover, $PSRR^+$ is relatively high at GBW in the proposed amplifier, and even improves if the load capacitance is increased. Also, a slew rate improvement method is proposed that improves slew rate of the amplifier by a factor of 2.38. In addition, simple and intuitive relationships are presented which fairly approximate locations of poles and zeros. The rest of this paper is organized as follows. Operation principles of the circuit are described in Section 2. Section 3 presents simulation results which demonstrate the circuit performance from various aspects, and also at different process corners and in presence of mismatch. Comparisons with other works in the literature are presented in Section 4. Finally, conclusions are presented in Section 5.

## 2 Proposed two-stage amplifier

### *2.1 Frequency compensation scheme*

Fig. 1*a* shows the conceptual diagram of the proposed amplifier.

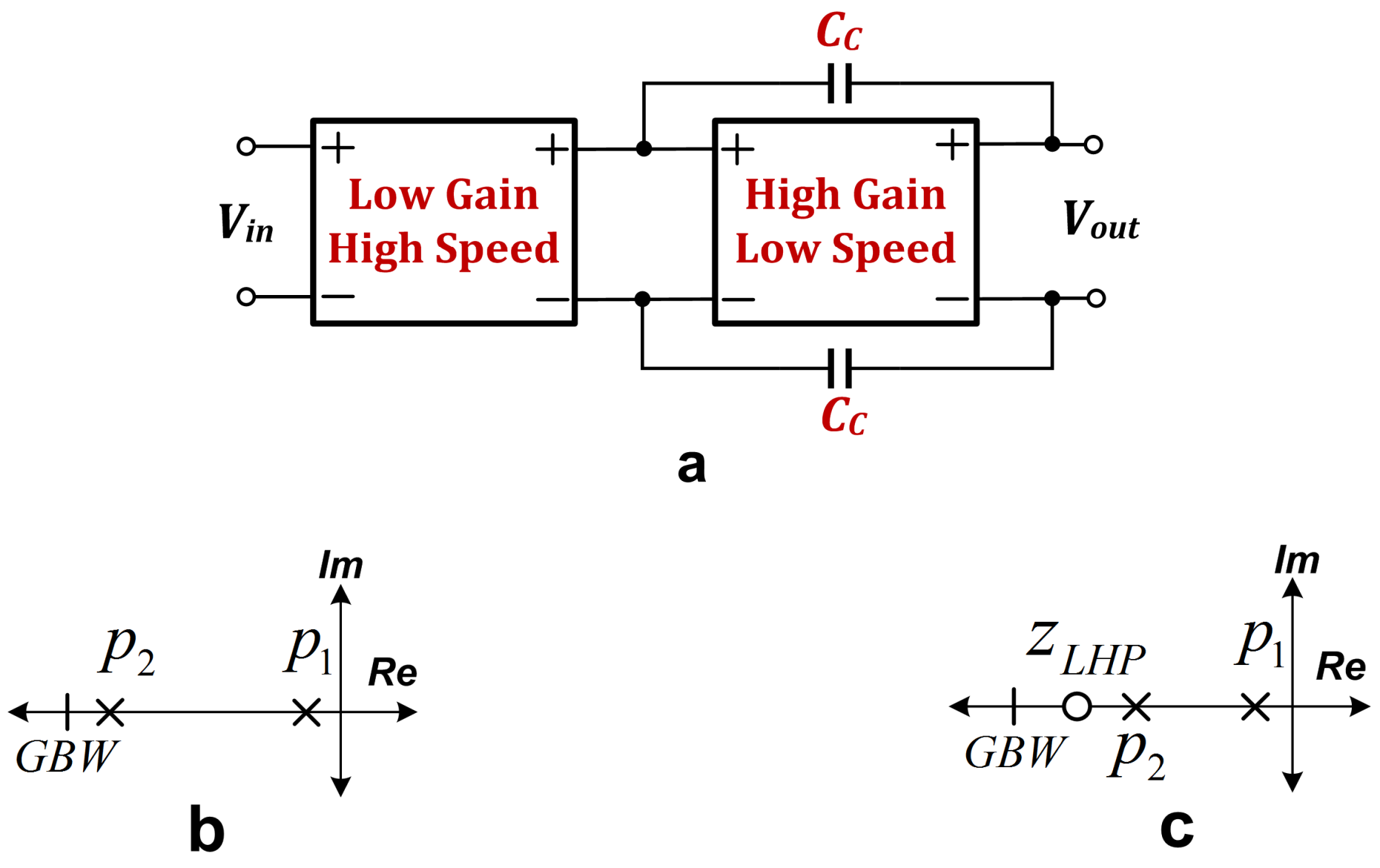

**Fig. 1** *basic idea and compensation scheme*

*a* Conceptual representation of the proposed amplifier

*b* Pole-zero locations of uncompensated proposed amplifier

*c* Pole-zero locations of compensated proposed amplifier

In the proposed structure, the first stage has lower gain, higher power consumption and consequently higher speed than the second stage. Therefore the dominant pole happens at the output of the second stage and the non-dominant one at the output of the first stage. A method similar to [20] is used in the proposed two-stage amplifier for compensation. The compensation capacitors, $C_C$, are placed between the inputs and outputs of the same polarity of the second stage. This introduces a zero in the LHP. By selecting the right value for this zero, we can eliminate effect of the non-dominant pole at output of the first stage. Consequently the output pole will be the dominant pole of the system. Figs. 1*b* and 1*c* show the pole-zero location of the amplifier before and after compensation, respectively. $p_1$, $p_2$, and $z_{LHP}$ indicate output pole of the second stage, output pole of the first stage and the LHP zero introduced by the compensation capacitors, respectively.

It is worth to note that the circuit is prone to instability due to use of positive feedback applied by the compensation capacitors. To prevent this problem, measures have been taken which will be explained in Section 2.6.

## 2.2 *Circuit description*

Fig. 2 shows the proposed two-stage amplifier. Devices $M_1$-$M_4$ and also $M_{t1}$ constitute the first stage, and $M_5$-$M_{10}$ and also $M_{t2}$ constitute the second stage. It is worth to mention that all NMOS and PMOS devices have their bulks connected to GND and $V_{DD}$, respectively. PMOS devices are used as input transistors which introduce less flicker noise than NMOS devices. Moreover, using PMOS devices at input, allows lower input common-mode (CM) level. Therefore, we can use NMOS transistors with smaller sizes as input switches in SC circuits and also reduce clock feed-through mismatch [5, 9, 21].

If the second stage introduces a large capacitance to nodes $N_1$ and $N_2$, pole of the first stage will be brought to lower frequencies and PM will be degraded. However, the cascode structure in the second stage reduces the Miller effect of gate-drain capacitance of $M_5$ and $M_6$ on nodes $N_1$ and $N_2$, respectively. Therefore, a smaller compensation capacitor, $C_C$, is required to compensate this effect.

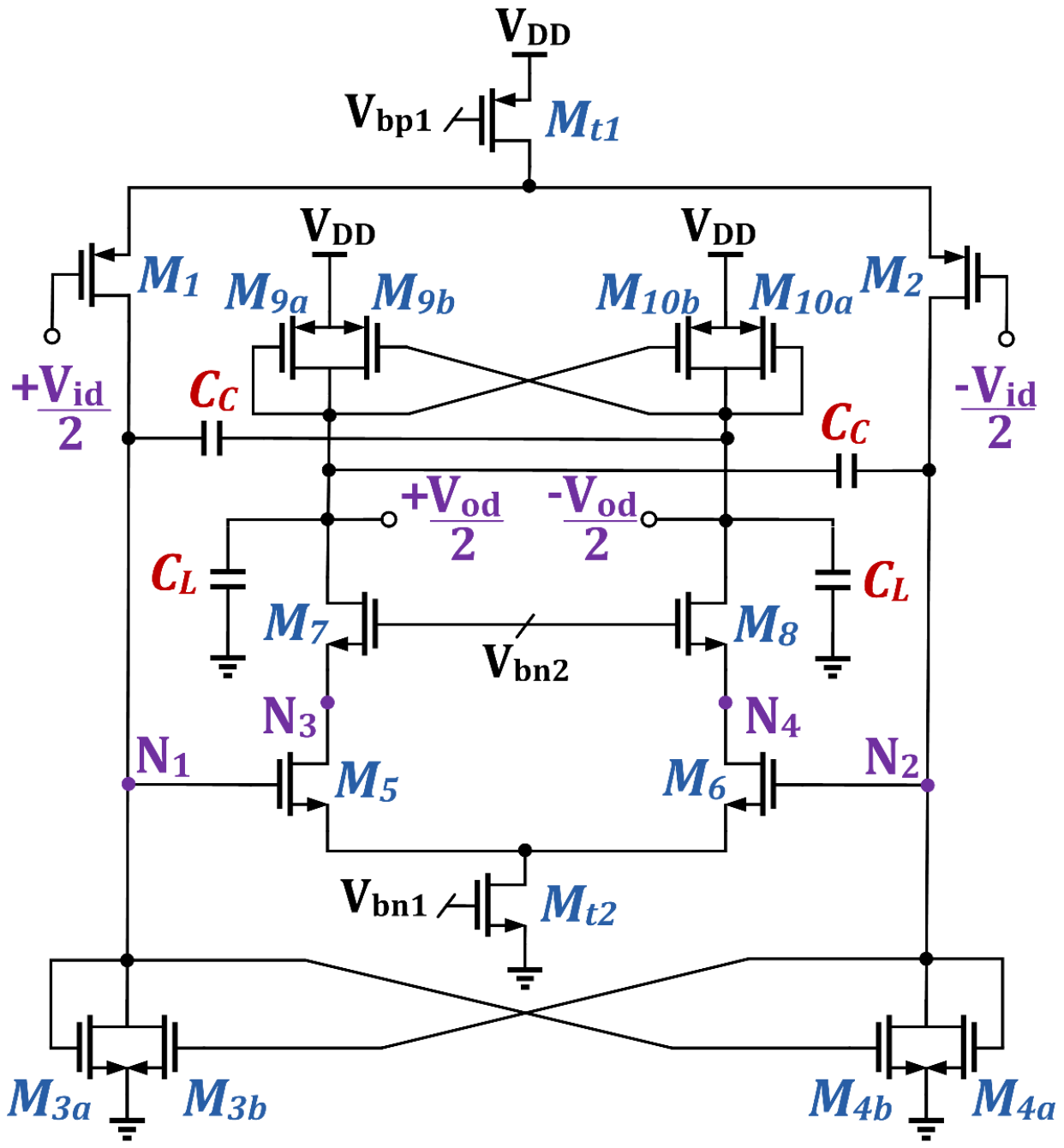

**Fig. 2** *Proposed two-stage amplifier*

Moreover, we can increase the output swing by choosing lower overdrive voltage for transistors of the second stage. At output of both stages, fully differential active loads have been used. The diode connected devices, $M_{3a}$ and $M_{4a}$ ($M_{9a}$ and $M_{10a}$), define the CM level conveniently, while $M_{3b}$ and $M_{4b}$ ($M_{9b}$ and $M_{10b}$) provide the desired output resistance by means of the positive feedback they apply to the output nodes. Therefore, no common-mode feedback (CMFB) circuit is required when this topology is used.

Another important feature of this topology is its higher CMRR than conventional fully differential amplifiers that use active load with CMFB circuit. In Fig. 3, $r_{Oi}$ indicates output resistance of a transistor, where i is index of transistors. In Fig. 3*a*, $\Delta V_x = -\Delta V_y$ for differential mode (DM) and if all transistors have the same size, then $g_{m3a}V_{gs3a} = -g_{m3b}V_{gs3b}$. In this case, the resistance seen from output nodes looking down is $r_{O3a} || r_{O3b} = r_{O3a}/2$ (Fig. 3*b*). For a CM signal $\Delta V_x = \Delta V_y$ and consequently $g_{m3a}V_{gs3a} = g_{m3b}V_{gs3b}$. The output resistance is simply $(1/2g_{m3a}) || (r_{O3a}/2)$ or approximately $1/2g_{m3a}$ for a CM signal (Fig. 3*c*), which is different from the one for DM. Therefore, output resistance and consequently gain is lower for CM than DM which means a higher CMRR.

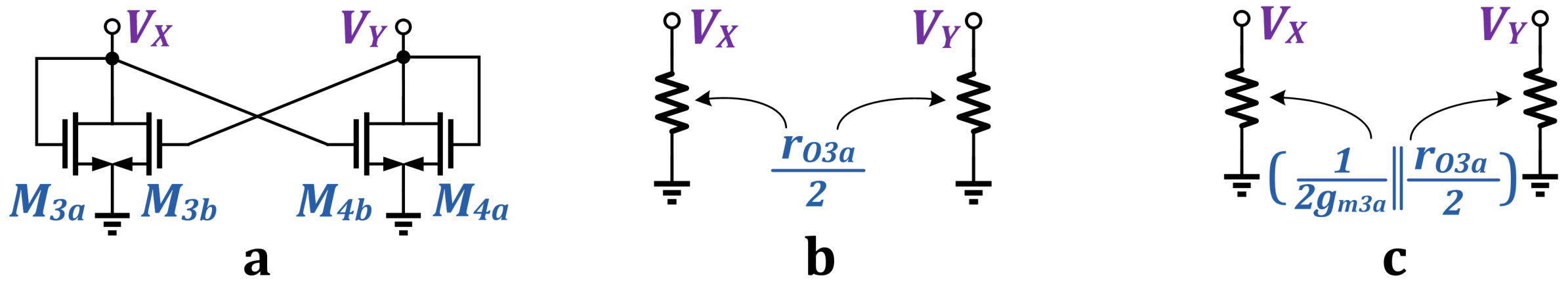

**Fig. 3** *Active load*

*a* Fully differential active load with positive feedback

*b* Equivalent resistance for DM analysis

*c* Equivalent resistance for CM analysis

Both stages show low sensitivity to process variations, which means the amplifier has almost equal performance at different process corners. Regarding the first stage, we should note that the input CM level is fixed, but threshold voltage of $M_1$ and $M_2$ varies at different process corners. This in turn changes drain-source voltage of $M_{t1}$ and causes its current to change due to channel length modulation. An increase in the bias current, increases the input transconductance and reduces the output resistance. Therefore, the gain experiences two opposing changes and consequently stays relatively fixed. However, input CM level of the second stage is determined by the diode connected transistors, $M_{3a}$ and $M_{4a}$. Consequently, neglecting variations of bias current of the first stage, CM level of $N_1$ and $N_2$ changes relative to the amount of threshold voltage variations in process corners. Therefore, unlike the first stage, in the second stage, drain-source voltage of $M_{t2}$ does not change in process corners, hence bias current of this stage stays fixed.

Mismatch between identically designed transistors can affect circuit performance. As mentioned before, if all devices in Fig. 3*a* have the same size, the positive transconductance introduced by $M_{3a}$ ($M_{4a}$) will cancel the negative transconductance introduced by $M_{3b}$ ($M_{4b}$), and the equivalent resistance of each node will be $(1/g_{ds3a})\|(1/g_{ds3b})=1/2g_{ds3a}$. However, in presence of mismatch, output resistance of the first stage will generally be:

$$R_1 = \frac{1}{g_{ds1} + g_{ds3a} + g_{ds3b} + g_{m3a} - g_{m3b}} \tag{1}$$

If $g_{m3b}$ becomes much larger than $g_{m3a}$ due to mismatch, it can cancel the other terms in denominator of (1). Consequently, $R_1$ will be negative and the circuit will be unstable. This means that the output will latch in one direction and will remain in that position regardless of the input.

Similarly, we can write for the output resistance of the second stage:

$$R_2 = \frac{1}{\left(1 \Big/ \left[ g_{ds5}^{-1} \left(1 + \left(g_{m7} + g_{mb7}\right) \times g_{ds7}^{-1}\right) + g_{ds7}^{-1} \right]\right) + g_{ds9a} + g_{ds9b} + g_{m9a} - g_{m9b}} \qquad (2)$$

Choosing $M_b$ to be conveniently smaller than $M_a$ can guarantee (1) and (2) to be positive, but results in lower output resistances, thus lower DC gain. We have chosen same sizes for devices $M_a$ and $M_b$ for the sake of better layout and lower parasitic capacitances. Over drive voltage of differential active load transistors in the first and second stages are designed to be quite high (200mV and 640mV, respectively). These choices cause active load transistors to have lower transconductance, therefore to be less prone to unstable operation. Simulations show that variations of $R_1$ and $R_2$ in presence of mismatch are reasonable, (1) and (2) always stay positive, and there is no chance of instability.

## 2.3 *Differential-mode analysis*

Fig. 4a shows an equivalent half circuit for the circuit in Fig. 2 and we use it to obtain differential gain ($V_{od}/V_{id}$).

Note that in Fig. 2 body effect is absent because sources of $M_1$ and $M_2$, and also $M_5$ and $M_6$, are virtual ground for DM operation. We can express the differential gain as:

$$A_d = \frac{V_{od}}{V_{id}} = -g_{m1} \times \left( \frac{r_{O3a}}{2} \,||\, r_{O1} \right) \times -g_{m5} \times \left( \frac{r_{O9a}}{2} \,||\, \left( r_{O7} + r_{O5} \left(1 + \left(g_{m7} + g_{mb7}\right) r_{O7}\right)\right)\right) \qquad (3)$$

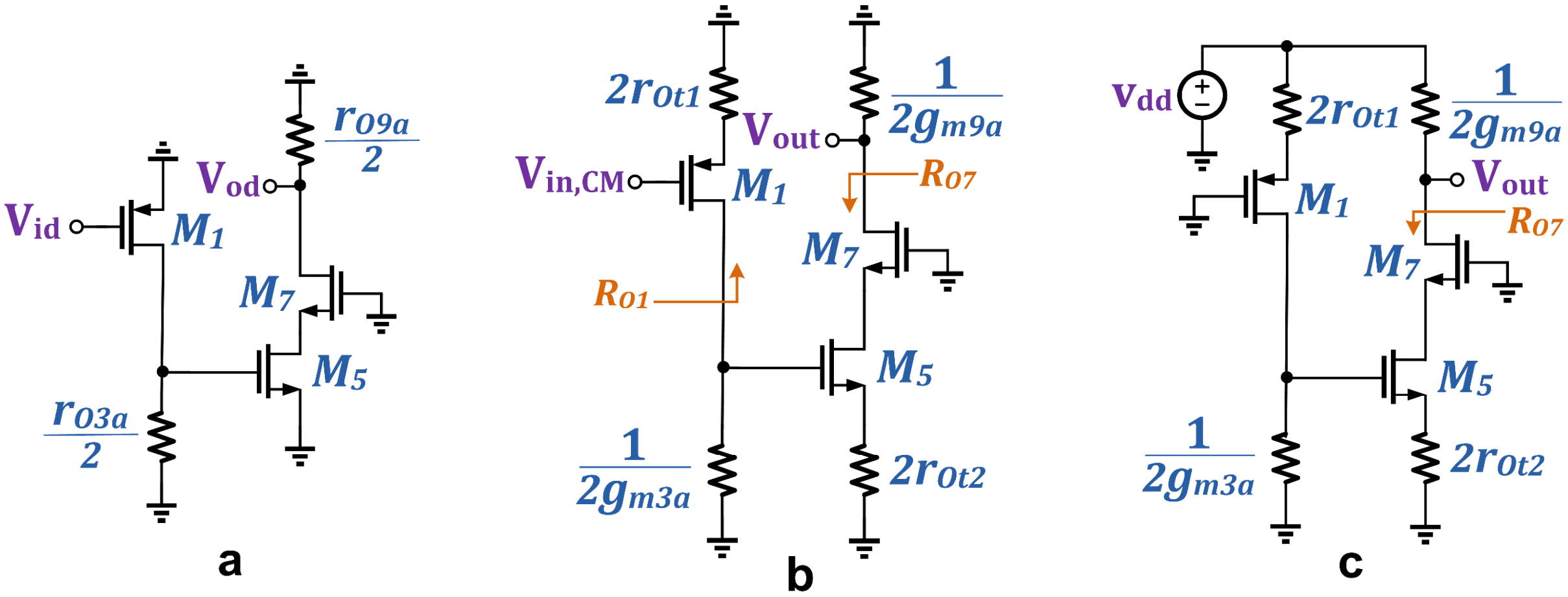

**Fig. 4** *Equivalent half circuit of Fig. 2 for DM, CM, and PSRR⁺ analysis*

*a* Equivalent half circuit for DM analysis

*b* Equivalent half circuit for CM analysis

*c* Equivalent half circuit for PSRR⁺ analysis

## *2.4 Common-mode analysis*

Fig. 4b shows the equivalent half circuit for CM analysis. For a common source stage with source degeneration $R_S$, including channel length modulation and body effect, we can obtain the transconductance $G_m$ as [9]:

$$G_m = \frac{g_m}{1 + \frac{R_S}{r_O} + R_S(g_m + g_{mb})} \tag{4}$$

Considering that $R_{O1}$ and $R_{O7}$ are much larger than $1/2g_{m3a}$ and $1/2g_{m9a}$ , and according to (4), we can express CM gain as:

$$A_{v,CM} \approx \frac{g_{m1}/2g_{m3a}}{1 + \frac{2r_{Ot1}}{r_{O1}} + 2r_{Ot1}(g_{m1} + g_{mb1})} \times \frac{g_{m5}/2g_{m9a}}{1 + \frac{2r_{Ot2}}{r_{O5}} + 2r_{Ot2}(g_{m5} + g_{mb5})} \tag{5}$$

The CMRR is defined as $CMRR = A_d/Av_{,CM}$, where $A_d$ is the DM gain, see(3), and $Av_{,CM}$ is the CM gain, see(5). As mentioned in Section 2.2, the proposed structure reduces $A_{v,CM}$ and delivers a higher CMRR.

## 2.5 *Positive power supply rejection ratio (PSRR⁺)*

Fig. 4c is used to evaluate effect of power supply voltage variations on output nodes. The voltage source $V_{dd}$ models the power supply variations. We can express the output voltage in terms of $V_{dd}$ as:

$$V_{out} = \left(V_{dd} \times -A_{v,CM}\right) + \left(V_{dd} \times \frac{R_{O7}}{R_{O7} + \left(1/2g_{m9a}\right)}\right) \approx V_{dd} \tag{6}$$

The first term in (6) represents the path through the first stage and then the second stage to the output nodes. This path has a gain equal to CM gain obtained in Section 2.4. The second term represents the voltage division between $1/2g_{m9a}$ and the resistance seen from drain of $M_7$ in Fig.4c. Finally, we can express $PSRR^+$ as:

$$PSRR^+ = \frac{V_{od}/V_{id}}{V_{out}/V_{dd}} \approx \frac{V_{od}}{V_{id}} = A_d \tag{7}$$

## 2.6 *Frequency response and stability*

Fig. 5a shows the simplified small signal model of the proposed two-stage amplifier.

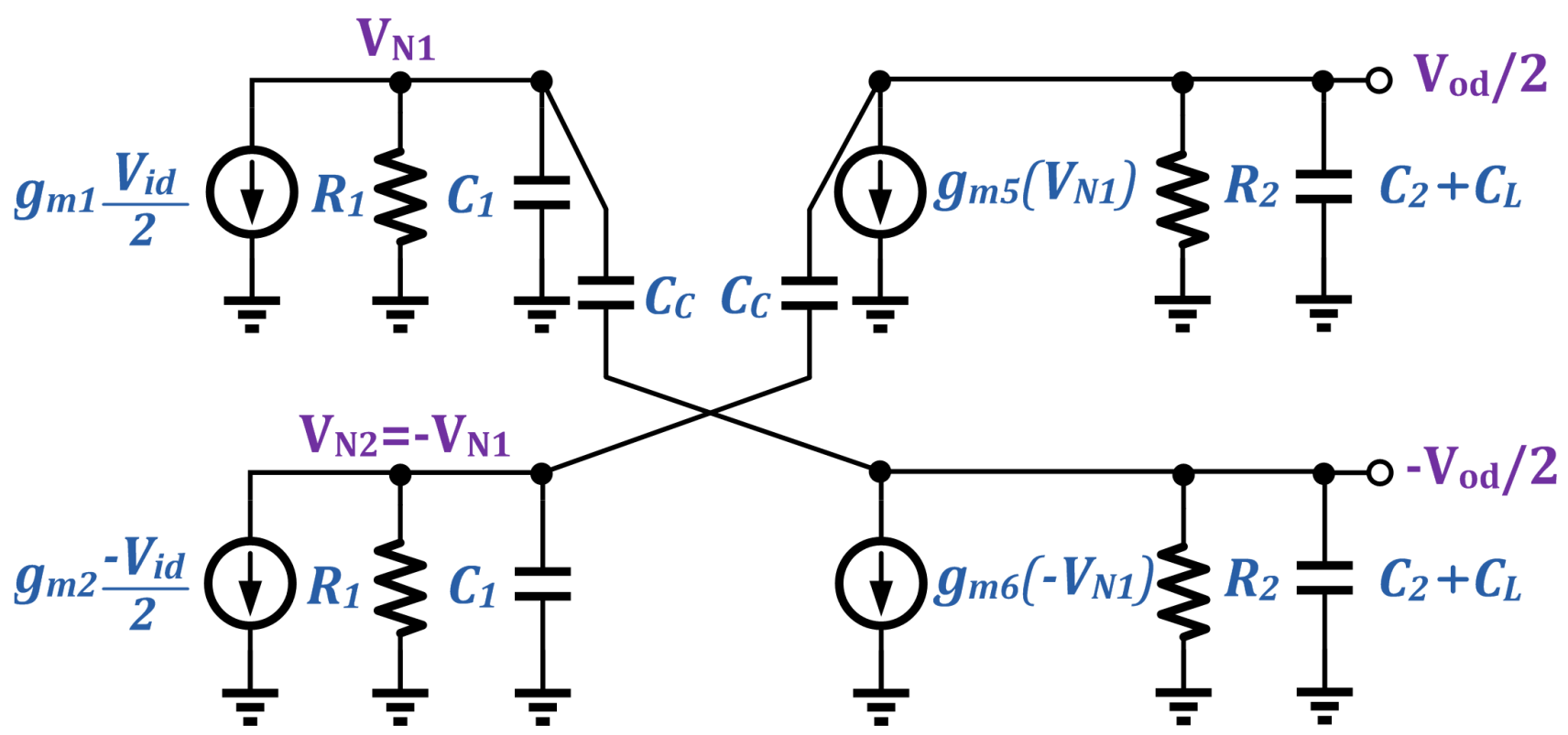

a

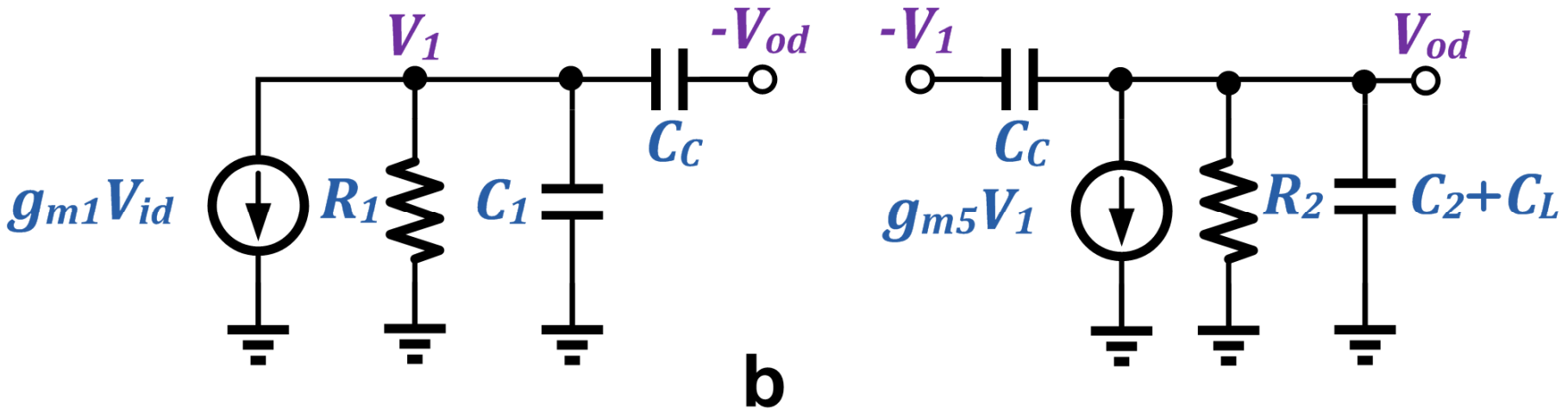

**Fig. 5** *Small-signal model for calculation of transfer function*

*a* Small-signal equivalent of the proposed amplifier

*b* Equivalent half circuit of the proposed amplifier

In this circuit, the second stage is considered as a first order system and effect of the cascode node has been neglected. This is completely safe, since frequency of the pole associated with the cascode node is much higher than other poles and has minimal effect on circuit performance.

The circuit shown in Fig. 5b is used to obtain the input-output transfer function, since poles of the differential circuit are the same as those of the equivalent half circuit. $R_1$ and $C_1$ are resistance and parasitic capacitance at output of the first stage, respectively. $R_2$ and $C_2$ indicate these parameters for the second stage.

According to Fig. 2, $C_1$ can be calculated as:

$$\begin{aligned} C_1 = {} & Cgs_{3a} + Cdb_{3a} + Cdb_{3b} + Cgd_{3b}\left(1 - \frac{1}{A_a}\right) \\ & + Cgs_{4b} + Cgd_{4b}\left(1 - A_a\right) + Cgs_5 \\ & + Cgd_5\left(1 - A_b\right) + Cdb_1 + Cgd_1\left(1 - \frac{1}{A_c}\right) \end{aligned} \tag{8}$$

Where $A_a = \frac{V_{N2}}{V_{N1}} = -1$, $A_b = \frac{V_{N3}}{V_{N1}} \approx -1$, and $A_c = \frac{V_{N1}}{V_{id}/2}$. NMOS devices of the first stage are identical.

Therefore (8) can be simplified to (9):

$$\begin{aligned} C_1 \approx {} & 2Cgs_{3a} + 2Cdb_{3a} + 4Cgd_{3b} \\ & + Cgs_5 + 2Cgd_5 + Cdb_1 + Cgd_1 \end{aligned} \tag{9}$$

Also, according to Fig. 2, $C_2$ can be calculated as:

$$C_2 \approx 2Cgs_{9a} + 2Cdb_{9a} + 4Cgd_{9b} + Cdb_7 + Cgd_7 \tag{10}$$

And finally, $R_1$ and $R_2$ can be expressed as:

$$R_1 = r_{O1} \parallel \frac{r_{O3a}}{2} \tag{11}$$

$$R_2 = \frac{r_{O9a}}{2} \parallel \left[ r_{O7} + r_{O5}\left(1 + \left(g_{m7} + g_{mb7}\right) r_{O7}\right) \right] \tag{12}$$

Writing KCL for nodes of the circuit in Fig. 5b gives:

$$g_{m1}V_{id} + \frac{V_1}{R_1} + C_1 S V_1 + C_C S\left(V_1 + V_{od}\right) = 0 \tag{13}$$

$$g_{m5}V_1 + \frac{V_{od}}{R_2} + \left(C_2 + C_L\right) S V_{od} + C_C S\left(V_{od} + V_1\right) = 0 \tag{14}$$

From (14), $V_1$ can be obtained as:

$$V_1 = -\frac{1 + \left(C_C + C_2 + C_L\right) R_2 S}{g_{m5} R_2 + C_C R_2 S} V_{od} \tag{15}$$

Substituting $V_1$ from (15) in (13), differential transfer function of the amplifier can be obtained as:

$$A_d(S) = \frac{V_{od}}{V_{id}}(S) = g_{m1} R_1 g_{m5} R_2 \frac{1 + \frac{C_C}{g_{m5}} S}{1 + \alpha S + \beta S^2} \tag{16}$$

Where $\alpha$ and $\beta$ are:

$$\alpha = R_2\left[C_2 + C_L + C_C\left(1 - g_{m5} R_1\right)\right] + \left(C_C + C_1\right) R_1 \tag{17}$$

$$\beta = R_1 R_2\left[\left(C_2 + C_L\right)\left(C_1 + C_C\right) + C_1 C_C\right] \tag{18}$$

Note that in (16), the system has a LHP zero at $\frac{g_{m5}}{C_C}$ which improves PM.

For a second order system, denominator of transfer function can be expressed as:

$$\begin{aligned} D(S) &= \left(1 - \frac{S}{p_1}\right)\left(1 - \frac{S}{p_2}\right) \\ &= 1 - \left(\frac{1}{p_1} + \frac{1}{p_2}\right) S + \frac{S^2}{p_1 p_2} \end{aligned} \tag{19}$$

Assuming that the two poles are well apart and $p_1$ is the dominant pole, (19) can be simplified to [10]:

$$D(S) \approx 1 - \frac{S}{p_1} + \frac{S^2}{p_1 p_2} \tag{20}$$

Comparing coefficients of $S$ in (20) and denominator of (16), the dominant pole is:

$$p_1 = \frac{-1}{R_2\left[C_2 + C_L + C_C\left(1 - g_{m5}R_1\right)\right] + \left(C_C + C_1\right)R_1} \tag{21}$$

In (21), $R_2$ is much larger than $R_1$ and $p_1$ can be simplified to (22), which is a good approximation for the dominant pole. This is because the second stage has a cascode structure and a bias current that is much lower than the first stage. Therefore, it has a much higher output resistance, hence $R_2$ is much larger than $R_1$.

$$p_1 \approx \frac{-1}{R_2\left[C_2 + C_L + C_C\left(1 - g_{m5}R_1\right)\right]} \tag{22}$$

For the amplifier to be stable, all poles should be located at LHP. This stability criterion suggests that:

$$C_2 + C_L + C_C\left(1 - g_{m5}R_1\right) > 0 \tag{23}$$

Which gives:

$$\frac{C_2 + C_L}{C_C} > g_{m5}R_1 - 1 \tag{24}$$

According to (24), increasing $C_L$ improves stability of the amplifier, similar to the case of a single pole structure. This is in contrast to conventional two-stage amplifiers, in which increasing the load degrades stability [9-11]. For the non-dominant pole we can write:

$$\begin{aligned} p_2 &= -\frac{R_2\left[(C_C + C_2 + C_L) - C_C g_{m5} R_1\right] + (C_C + C_1)R_1}{R_1 R_2\left[\left(C_2 + C_L\right)(C_1 + C_C) + C_1 C_C\right]} \\ &\approx \frac{-1}{R_1\left[\dfrac{\left(C_2 + C_L\right)(C_1 + C_C) + C_1 C_C}{C_2 + C_L + C_C(1 - g_{m5}R_1)}\right]} \end{aligned} \tag{25}$$

Note that (22) suggests that the frequency compensation method increases bandwidth simply if $g_{m5}R_1>1$. In fact both $g_{m5}$ and $R_1$ can be used to increase bandwidth and also DC gain, see(16). It is worth to note that if we use $g_{m5}$ to increase bandwidth, the LHP zero will be pushed to higher frequencies and PM will be reduced. In this case we can increase the compensation capacitor to move the zero to lower frequencies to compensate for the reduction of PM. According to (22), increasing $C_C$ increases bandwidth for $g_{m5}R_1>1$ . If we increase $R_1$ to increase the bandwidth, (25) suggests that $p_2$ will be pushed to lower frequencies and PM will be reduced. Similar to the case of using $g_{m5}$, we can compensate for this reduction by increasing the compensation capacitor. Therefore, (16), (22), and (25) provide us with insight on how to employ different parameters to achieve the desired specifications for the circuit.

## 2.7 *Slew rate enhancement*

Fig. 6 shows the proposed amplifier with the slew rate enhancement circuit which is composed of $M_{S1}$ and $M_{S2}$. In normal condition, $M_{S1}$ and $M_{S2}$ operate in sub threshold region and sink a small current. When a positive input is applied, voltage of node $N_1$ drops which turns $M_5$ and $M_7$ off and increases voltage of node $N_3$. Therefore, $M_{S1}$ turns on and sinks current from node $-V_{od}/2$ which discharges $C_L$ faster. This is the intended SR enhancement. $M_{S2}$ plays the same role for node $V_{od}/2$ when a negative input is applied. Connecting $M_{S1}$ and $M_{S2}$ to nodes $N_3$ and $N_4$ adds to the parasitic capacitance of these nodes, thus moves their poles to lower frequencies and degrades the PM. However, this can be easily compensated by a small increase in the compensation capacitors $C_C$.

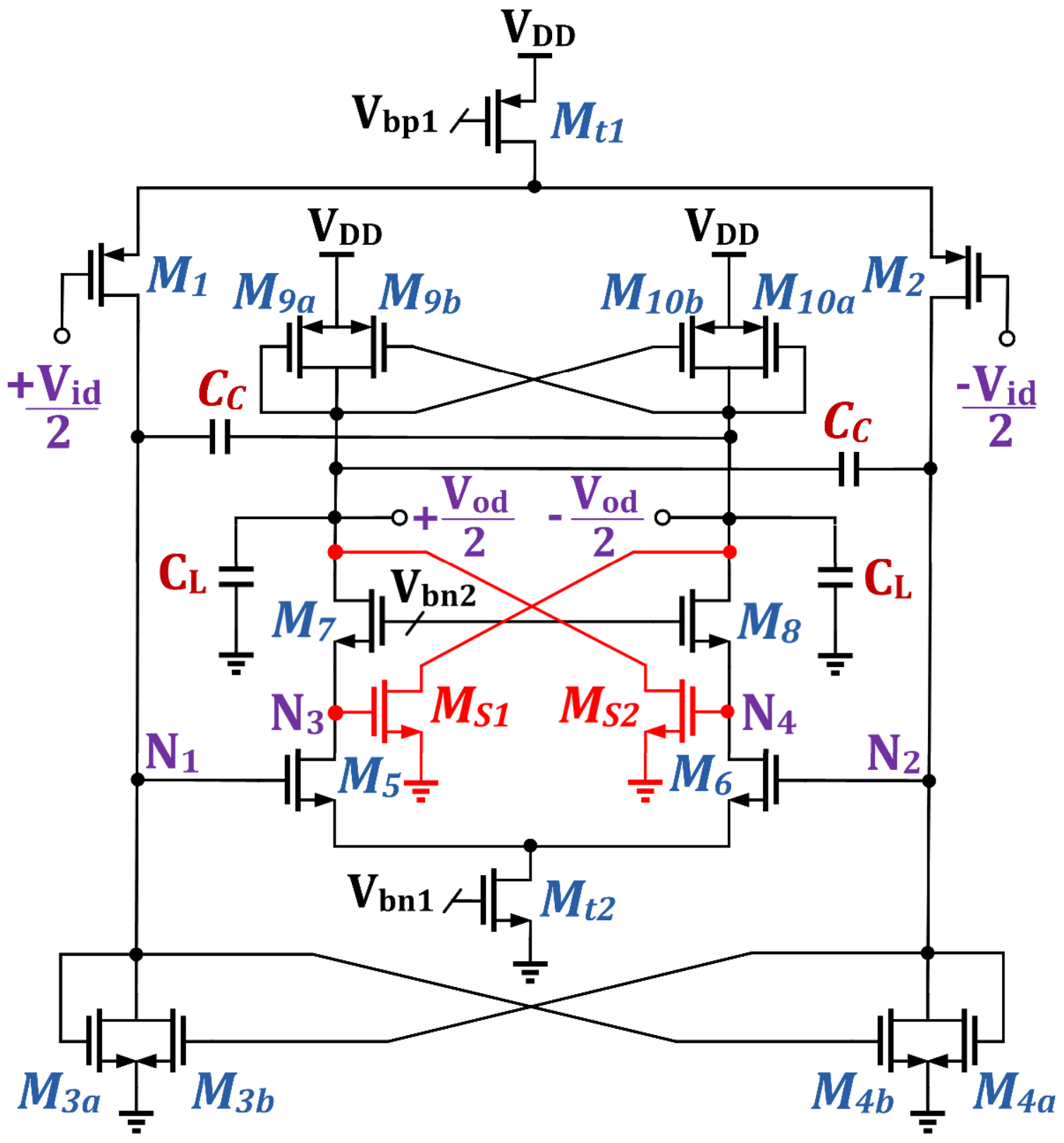

**Fig. 6** *The proposed amplifier with slew enhancement circuitry*

## 3 Simulation results

All simulations presented in this article have been performed in HSPICE simulator using level 49 models for 0.18μm CMOS process. The whole circuit consumes approximately 0.86mW from a 1.8V supply voltage. The compensation capacitors are 0.75 pF. Table 1 summarizes component sizes of the proposed two-stage amplifier.

**Table1** Amplifier device sizes

| **Devices** | ***W/L*** ($\mu m/\mu m$) | $g_m$ (ms) | $g_{mb}$ (ms) | $r_O$ ($k\Omega$) |
|---|---|---|---|---|
| $M_1, M_2$ | 200/0.18 | 4.22 | 1.17 | 10.95 |
| $M_{3a}, M_{4a}$ $M_{3b}, M_{4b}$ | 10/0.5 | 1 | 0 | 114.6 |
| $M_5, M_6$ | 5/0.18 | 0.21 | 0.049 | 66 |
| $M_7, M_8$ | 5/0.18 | 0.214 | 0.046 | 128.2 |
| $M_{9a}, M_{10a}$ $M_{9b}, M_{10b}$ | 2/3 | 0.018 | 0 | 12500 |
| $M_{t1}$ | 500/0.18 | - | 0 | 4.98 |
| $M_{t2}$ | 10/0.5 | - | 0 | 54.6 |
| $M_{S1}, M_{S2}$ | 50/0.18 | - | - | - |

Bias currents of the first and second stage are 456uA and 22uA, respectively. Fig. 7 shows the frequency response, PSRR$^+$, and CMRR for $C_L$=5pF. Low frequency open loop gain, GBW, and PM are 82.7dB, 88.9MHz and 68.7°, respectively.

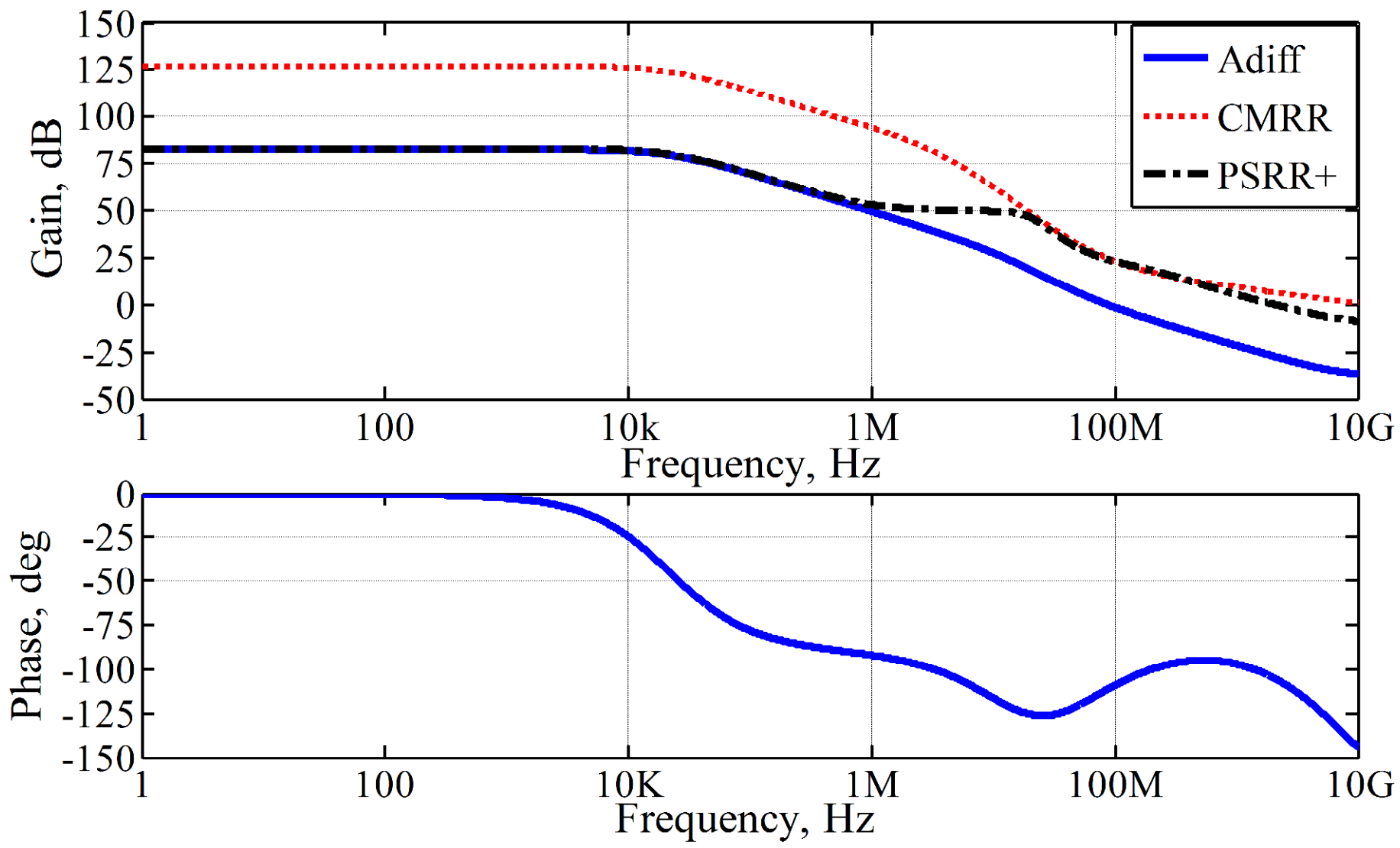

**Fig. 7** *Open loop frequency response of the proposed amplifier*

Contributions of the first and second stage to the open loop gain are 31.8dB and 50.9dB, respectively. Also the lowest PM is 53.7° and happens at 26.3MHZ with15.3 dB gain. $PSRR^+$ and CMRR at low frequencies are 83.2dB and 127dB, respectively. They are 24.2dB and 24.8dB at GBW, respectively. The -3dB bandwidth of $PSRR^+$ is 22KHz which is larger than that of the Miller compensated amplifiers [16, 17].

Table 2 provides a comparison between the estimated values for some of the amplifier parameters and their spice results and indicates a good agreement between the two. Values for $C_1$ and $C_2$ as defined in (9) and (10) are obtained from operating point analysis in spice and are 325fF and 137fF, respectively.

**Table 2** Comparison between the estimated and spice values

| Description | Estimated | Spice |
|---|---|---|
| **$A_V$ [dB]** | 82.98 | 82.7 |
| **$p_1$[Mrad/s]** | 0.130 | 0.139 |
| **$p_2$[Mrad/s]** | 83.855 | 81.98 |
| **$z_{LHP}$[Mrad/s]** | 277.72 | 321.69 |
| **CMRR[dB]** | 125.9 | 127 |
| **$PSRR^+$[dB]** | 82.98 | 83.2 |

Table 3 shows effect of $C_L$ on some of the circuit parameters. Note that unlike the conventional two-stage amplifiers in which $PSRR^+$ approaches 0dB around the GBW, it is more than 24dB in the proposed amplifier.

**Table 3** Effect of CL on some of performance parameters

| $C_L$ [pF] | GBW [MHz] | PM [deg] | CMRR@ GBW[dB] | $PSRR^+$@ GBW [dB] |
|---|---|---|---|---|
| **10** | 53.8 | 61.7 | 31 | 29.1 |
| **20** | 34 | 57.9 | 38.3 | 36.4 |
| **30** | 26.2 | 57.8 | 43 | 41.3 |
| **40** | 21.7 | 58.9 | 46.5 | 44.4 |
| **50** | 18.7 | 60.3 | 49.4 | 46.2 |

Table 4 presents frequency characteristics of the proposed amplifier at different process corners for $C_L$=5 pF. We have set the compensation capacitors to 0.75pF and swept the load capacitance from 5pF to 100pF. The lowest PM is 46.9° which happens at FF corner, for $C_L$=20pF and at GBW of 44.8MHz. If we further increase the load capacitance beyond 100pF, the system keeps stable and acts similar to a single pole system.

**Table 4** Performance at process corners

| Process corners | $A_V$ [dB] | GBW [MHz] | PM [deg] |
|---|---|---|---|
| **FF** | 82.3 | 103 | 55.9 |
| **FS** | 84.1 | 89.6 | 61.8 |
| **SF** | 82.8 | 89.9 | 71.7 |
| **SS** | 83.9 | 82.3 | 69.2 |

Fig. 8a shows the proposed amplifier in a unity gain configuration, where $C_C$=0.75pF, $C$=1pF, $R$= 5Meg, and $C_L$= 5pF.

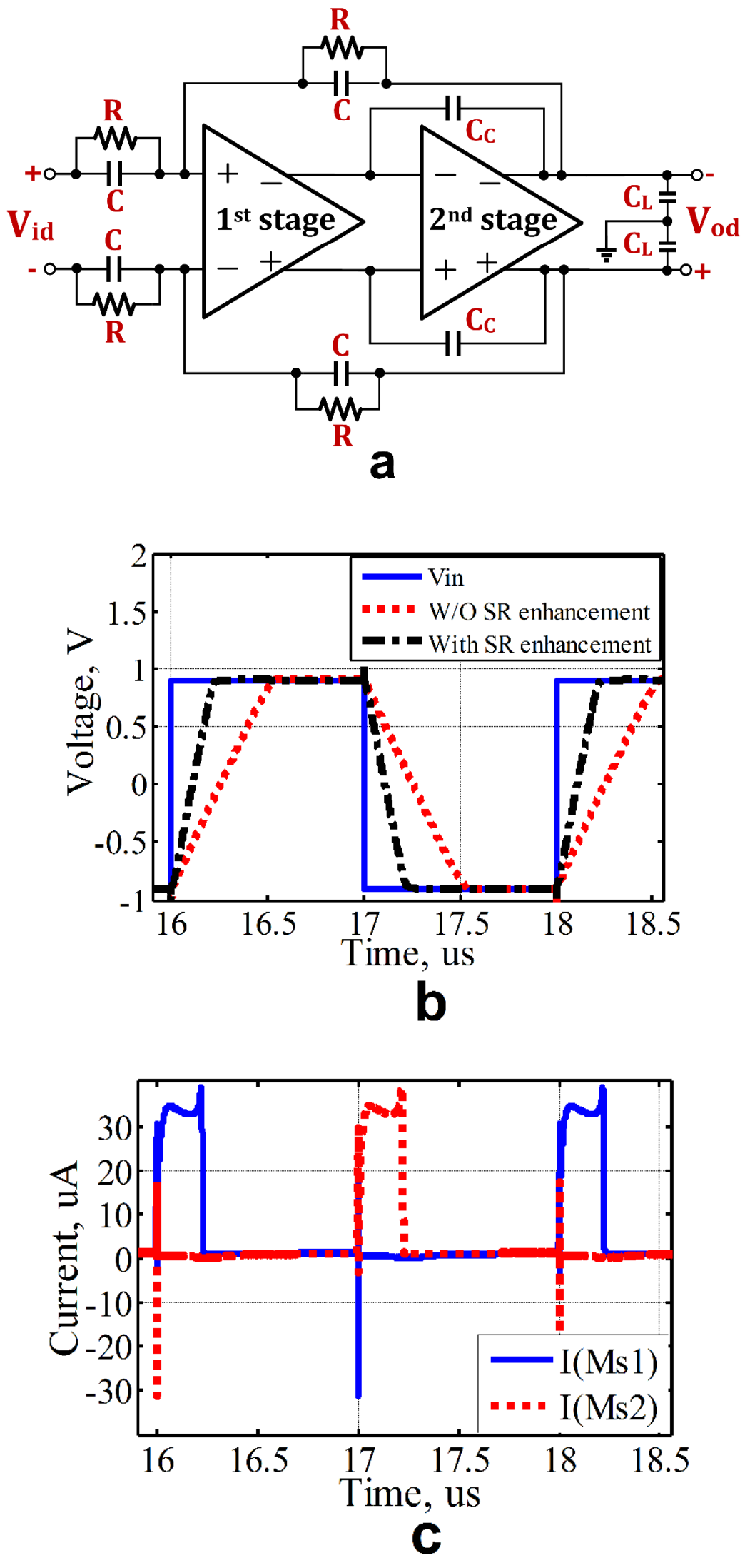

**Fig. 8** *Closed loop simulation of the proposed amplifier*

*a* Unity gain amplifier

*b* Transient response of the proposed amplifier at 500KHz, 1.8$V_{pp}$ step

*c* Currents of slew rate enhancement transistors

Fig. 8b shows the step response of the unity gain amplifier with 1.8$V_{p\text{-}p}$ output swing. Note that the proposed method improves SR from 3.55(V/μs) to 8.67(V/μs). As shown in Fig. 8c, $M_{S1}$ and $M_{S2}$ sink a small current (less than 1uA) when the circuit is at rest. $M_{S1}$ and $M_{S2}$ only sink current at positive and negative peaks of the output, respectively. Also after adding the slew rate enhancement circuit, the 1% and 0.1% settling times for 0.5V input signal are 61.7ns and 79.5ns, respectively.

Fig. 9 shows THD of the proposed amplifier in process corners for different output swings, all at 500kHz input frequency. Note that the proposed circuit has better than -54dB linearity at all process corners for output swings up to 1.8$V_{p\text{-}p}$, which means less than 0.2% nonlinearity error.

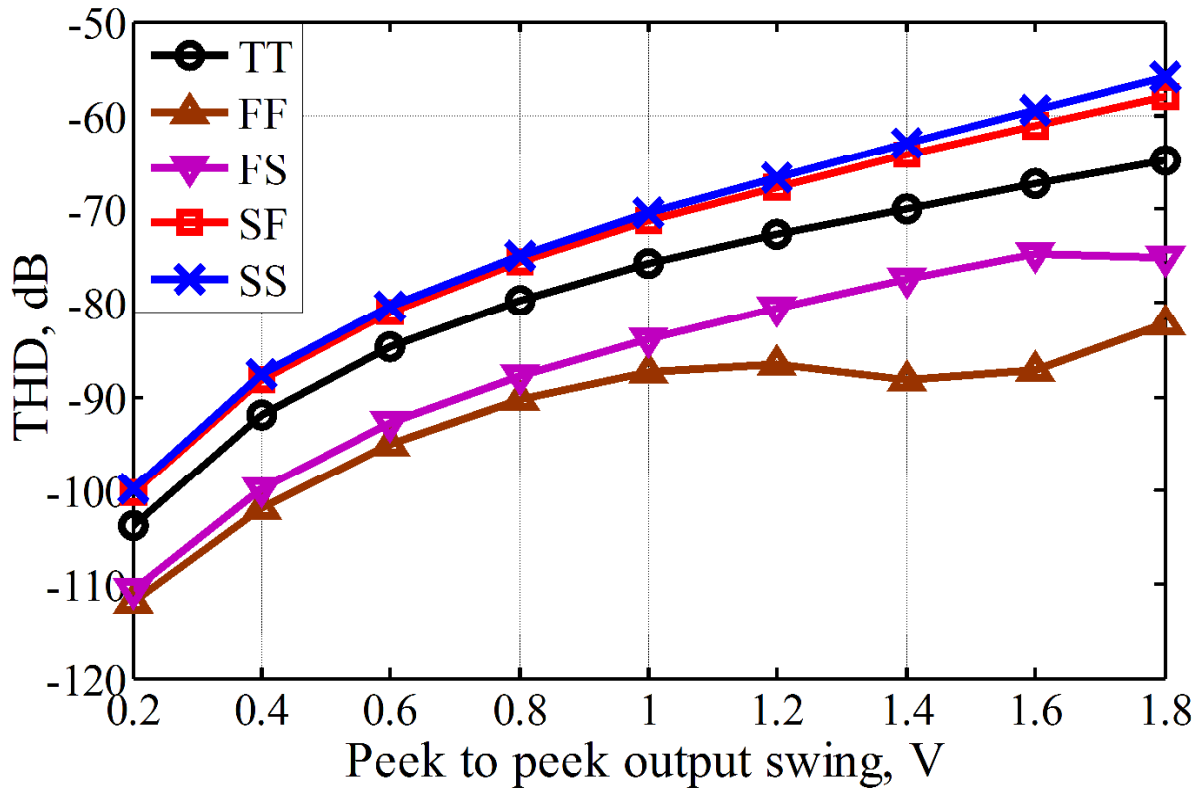

**Fig. 9** *THD versus output swing in unity gain configuration at different process corners*

Fig. 10 shows layout of the proposed amplifier which has an area of approximately 49.47μm×61.42μm. Post layout simulations show that GBW is reduced from 88.9MHz to 82.9MHz and PM is reduced from 68.7° to 67.3°, which are convenient.

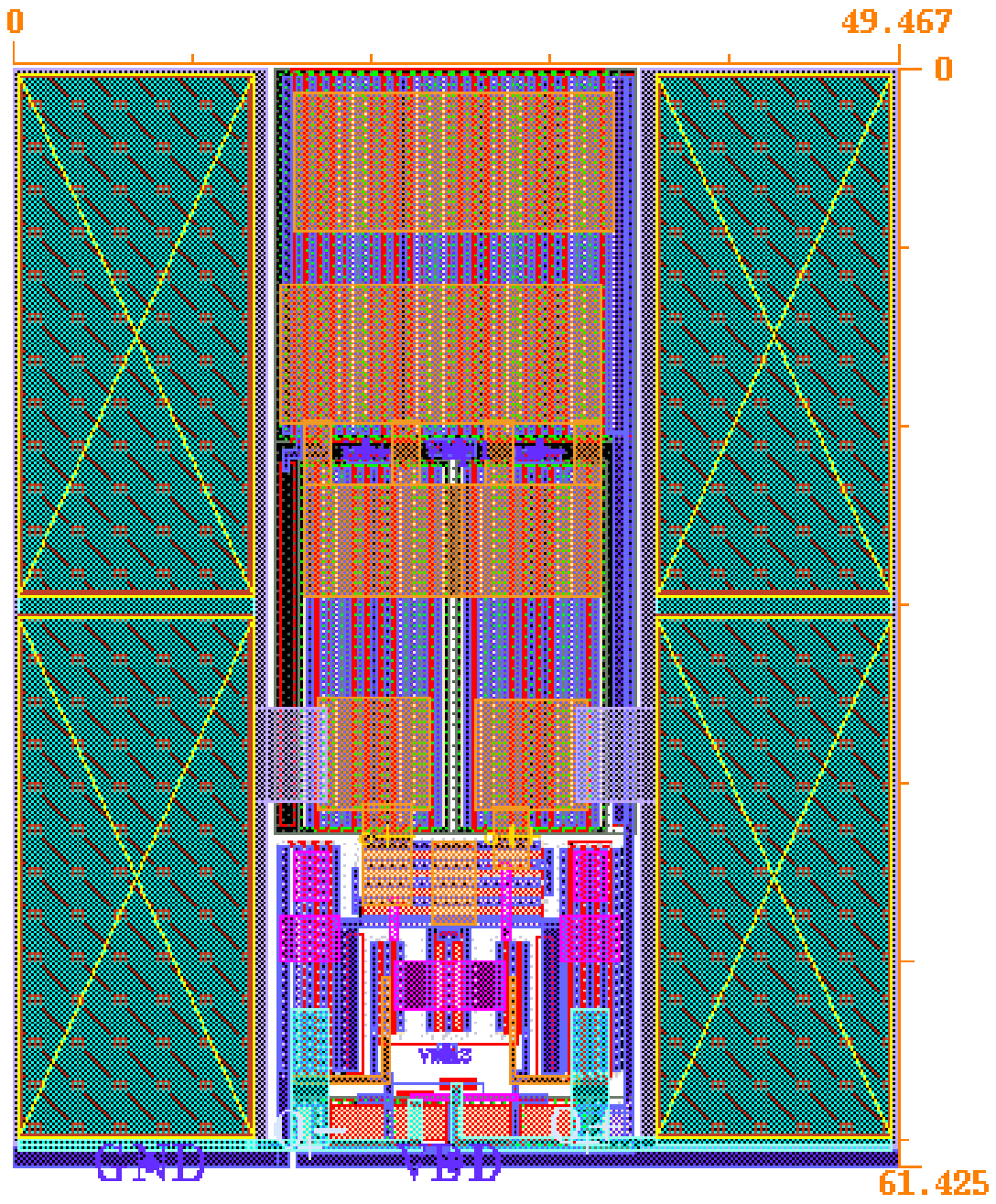

**Fig. 10** *layout of the proposed two-stage amplifier*

Devices fabricated in a CMOS process exhibit two components of mismatch: Systematic and Random (stochastic). Non-uniform thermal distribution during the fabrication process, dimensional errors, device orientation, etc, are some possible reasons for systematic mismatch. However, systematic mismatch is deterministic, predictable and can be extensively reduced with proper layout. Random mismatch represents the stochastic and unpredictable portion of mismatch. Statistical variations in the number of dopant atoms and dopant diffusion, edge roughness, polysilicon grain effects, etc, are some possible reasons for random mismatch.

In our design, systematic mismatch can be greatly reduced with proper layout and implementing transistors close to each other. Therefore, there remains only the random mismatch. Let the current in an MOS transistor to be:

$$I_D = \frac{1}{2}\beta(V_{GS} - V_{TH})^2 \tag{26}$$

Where $\beta = \mu_n \cdot C_{OX} \cdot \frac{W}{L}$. For a pair of MOS transistors in close proximity, it can be shown that [9, 22, 23]

$$\sigma(V_t) = \frac{A_{V_t}}{\sqrt{WL}} \tag{27}$$

$$\sigma(\beta) = \frac{A_\beta}{\sqrt{WL}} \tag{28}$$

Where $\sigma(V_t)$ is the standard deviation of the threshold voltage, $\sigma(\beta)$ is the standard deviation of the current factor, and $A_{V_t}$ and $A_\beta$ are technology dependent parameters. W and L are width and channel length of transistors, respectively.

Monte Carlo analysis has been performed to evaluate the circuit performance in presence of mismatch. For the sake of safety, $A_{V_t}$ is chosen to be 6mv.um and 6.6mv.um for NMOS and PMOS devices, respectively, which are more pessimistic than the typical values provided in [22, 24] for 0.18µm CMOS process. Also, $A_\beta$ is chosen to be 1.04%.um and 0.99%.um for NMOS and PMOS devices, respectively.

Fig. 11 shows results of 30 Monte Carlo runs for the proposed amplifier at TT corner. It is worth to mention that for each run, the input referred offset is obtained and applied to the input, so that the DC offset at the output is zero. Maximum and minimum $A_V$ at different runs are 84.8dB and 80.7dB, respectively.

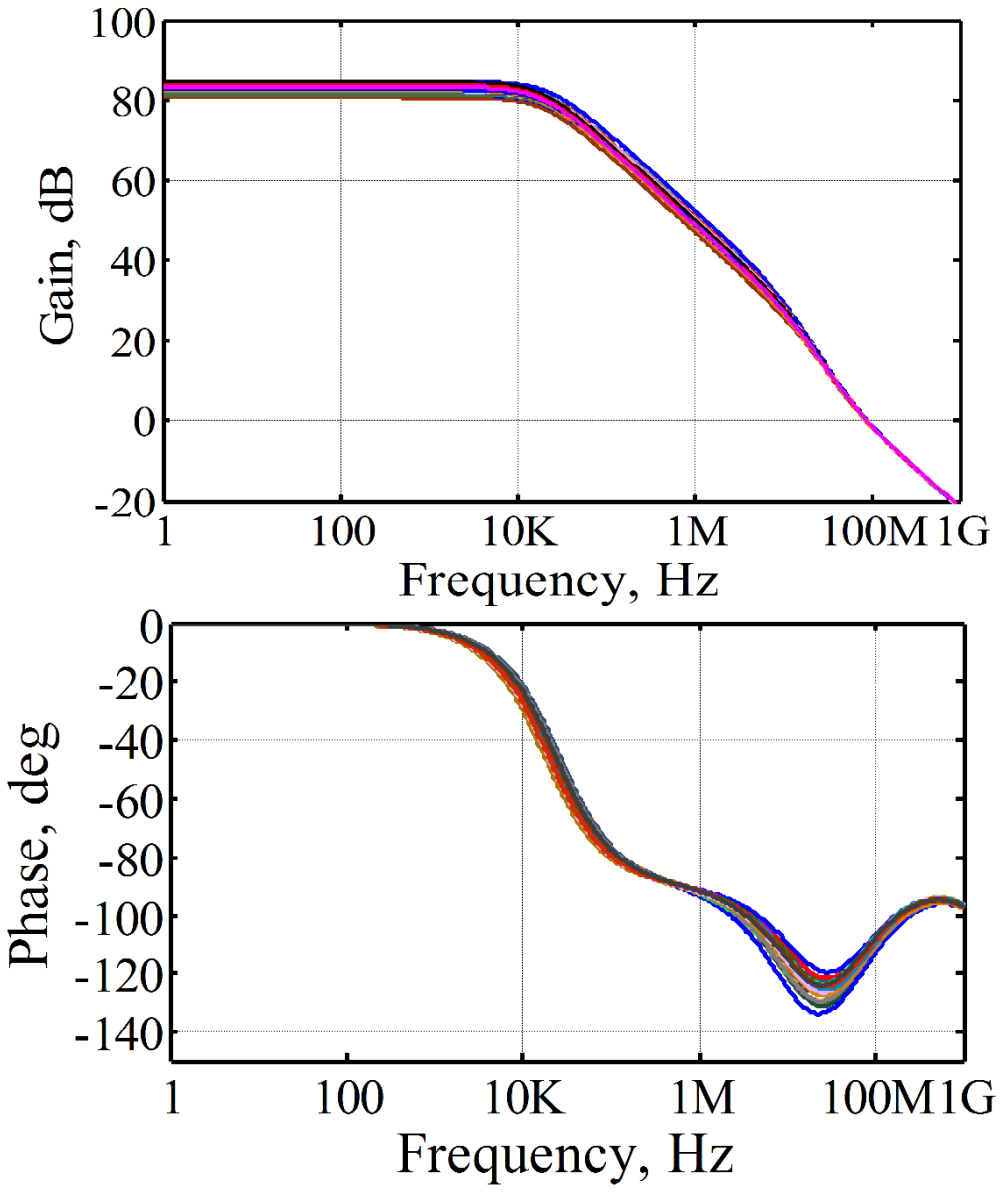

**Fig. 11** *Monte Carlo simulation for frequency response of the proposed amplifier*

Fig. 12 shows variations of resistance of the output nodes of both stages from (1) and (2) in Monte Carlo analysis. Fig. 12c shows product of *R1* and *R2*, whose largest variation from the nominal value is less than a factor of two. This shows that the open loop gain is not very sensitive to mismatch (less than 6dB) and agrees with results of Fig. 11.

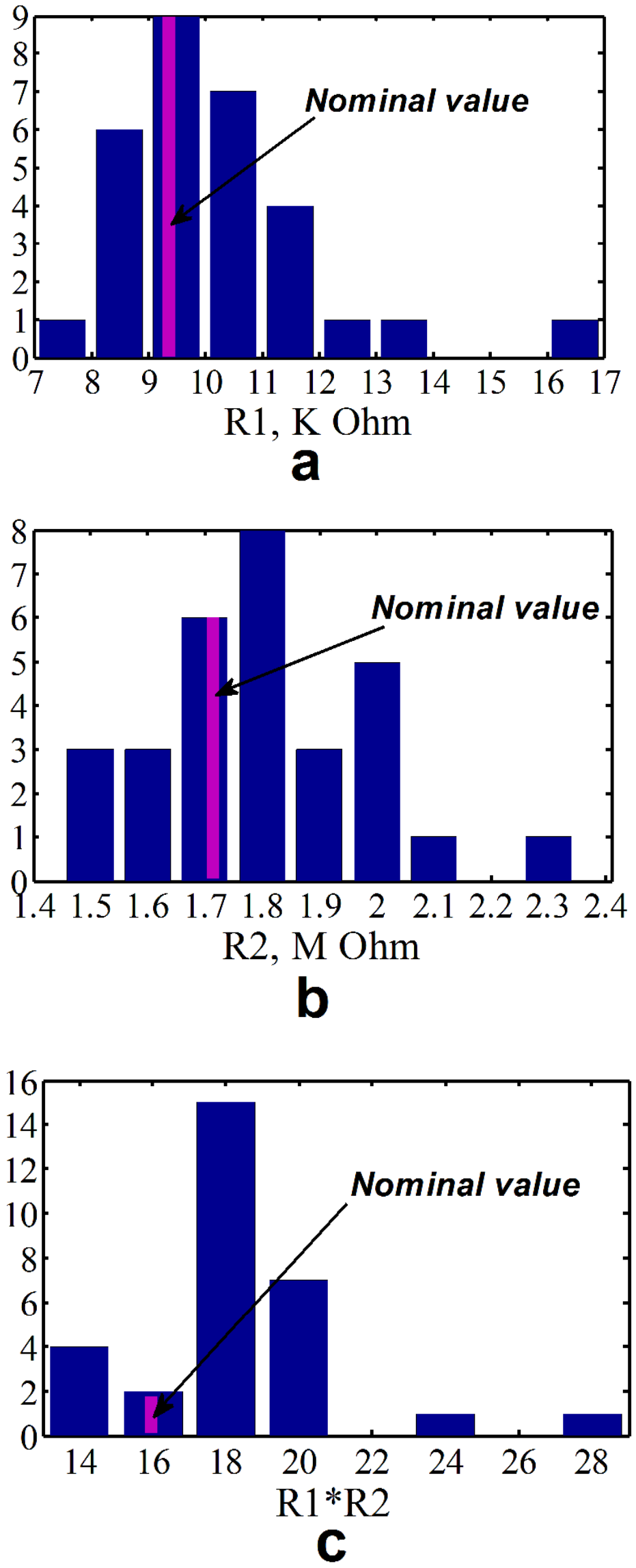

**Fig. 12** *Statistical distribution of R1 and R2 in 30 Monte Carlo runs.*

*a R1*
*b R2*
*c* Product of *R1* and *R2*

Table 5 lists results of Monte Carlo analysis at different process corners for $C_L$=5pF. The Nominal values correspond to the case of no mismatch, whereas Max and Min show the maximum and minimum values obtained from Monte Carlo analysis for GBW and Av. Table 5 shows that mismatch causes the largest variation in the open loop gain at FF corner, where it is increased by 3dB. Also, mismatch causes the largest variation in GBW at FS corner, where it is increased by 2.34%. At presence of mismatch, the lowest PM is 44°, and happens at FF corner at GBW=46.1MHz for $C_L$=20pF. Also, the largest input referred offset is 3.2mV and happens at FS corner.

**Table 5** Mismatch effect on some of performance parameters in different process corners

| Process Corners | $A_V$[dB] | | | GBW[MHz] | | |
|---|---|---|---|---|---|---|
| | Min | Nominal | Max | Min | Nominal | Max |
| **TT** | 80.7 | 82.7 | 84.8 | 86.9 | 88.9 | 90.9 |
| **FF** | 79.5 | 82.3 | 85.3 | 101 | 103 | 105 |
| **FS** | 81.3 | 84.1 | 86.2 | 87.8 | 89.6 | 91.7 |
| **SF** | 80.6 | 82.8 | 85 | 87.9 | 89.9 | 92 |
| **SS** | 81.7 | 83.9 | 86 | 80.4 | 82.3 | 84.2 |

Also, the maximum and minimum settling times (with 0.1% accuracy) at different Monte Carlo runs are 81ns and 67ns, for 0.5V input signal in a unity gain configuration and $C_L$=5pF. As simulation results show, the proposed amplifier has a good performance at presence of mismatch, which validates correct performance of the differential active load (Fig. 3), and its immunity to instability. Note that for the sake of safety, the standard deviation of the threshold voltage used in simulations, $\sigma_{vt}$, is chosen to be larger than those reported in the literature.

Table 6 summarizes main features of the proposed amplifier in Fig. 6.

**Table 6** Amplifier characterization results

| Description | | Features |
|---|---|---|
| **Technology [µm]** | | 0.18 |
| **Supply [V]** | | 1.8 |
| **Power Consumption [mW]** | | 0.86 |
| **Capacitive Load [pF]** | | 5 |
| **Compensation Capacitor [pF]** | | 0.75 |
| **DC Gain [dB]** | | 82.7 |
| **GBW [MHz]** | | 88.9 |
| **PM [deg]** | | 68.7 |
| **Slew Rate [V/µs]** | | 8.67 |
| **Area [$\mu m^2$]** | | 3038.5 |
| **Maximum Input offset [mV]** | | 3.2 |
| **$Vin_{p-p}$= 0.5V** | 1% settling time [ns] | 61.7 |
| | 0.1% settling time [ns] | 79.5 |
| **CMRR[dB]** | $F_{in}$=5KHz | 127 |
| | $F_{in}$=GBW | 24.8 |
| **$PSRR^+$[dB]** | $F_{in}$=5KHz | 83.2 |
| | $F_{in}$=GBW | 24.2 |
| **Input noise [$nV/Hz^{1/2}$]** | $F_{in}$=10KHz | 39.8 |
| | $F_{in}$=1MHz | 4.68 |

# 4 Comparison

Table 7 provides a comparison between the proposed amplifier and some of other single-stage, two-stage, and three-stage amplifiers reported in the literature.

**Table 7** Comparison table

| Parameter | | This work simulated | [5] measured | | [26] measured | [27] measured | [28] simulated |
|---|---|---|---|---|---|---|---|
| **Number of stages** | | 2 | 1 | | 2 | 2 | 2 |
| **Technology (µm)** | | 0.18 | 0.18 | | 0.065 | 0.13 | 0.065 |
| **Supply (V)** | | 1.8 | 1.8 | | 1 | 1 | 1.2 |
| $C_{Load}$ | | 5 | 5.6 | | 2 | >5.5 | 0.25 |
| **Power (mW)** | | 0.86 | 1.44 | 0.72 | 1.6 | 0.11 | 0.24 |
| **DC Gain** | **(V/V)** | 13700 | 478.6 | 556 | 640 | >3162 | 2290 |
| | **dB** | 82.7 | 53.6 | 54.9 | 56.1 | 70 | 67.2 |
| **GBW (MHz)** | | 88.7 | 134.2 | 70.4 | 450 | 35 | 321.5 |
| **PM (deg)** | | 68.7 | 70.6 | 79.8 | 77 | >45 | 61 |
| **Area (µm$^2$)** | | 3038.5 | 4958.2 | 3001.8 | 1800 | 12351 | N/A |
| **1% settling (ns) @ $Vin_{pp}$ (V)** | | 20.5 @ 0.1<br>61.7 @ 0.5 | 11.2@0.1 | 20.8@0.1 | 10 @ 0.5 | 134 @ 0.1 | N/A |
| **0.1% settling (ns) @ $Vin_{pp}$ (V)** | | 25.8 @ 0.1<br>79.5 @ 0.5 | N/A | | N/A | N/A | 18 @ 0.3 |
| **SR(V/µs)** | | 8.67 | 94.1 | 48.1 | N/A | 19.5 | 84.5 |
| **Input offset [mV]** | | 3.2 | 7.6 | 11.1 | N/A | N/A | N/A |

In [5], two designs have been presented based on the proposed topology there. Compared to the amplifier in [5], the proposed amplifier consumes less power and area, but has 28 times higher gain using the same technology. However, its slew rate is almost 11 times smaller than [5]. It is worth to note that in [5], we can reduce the bias current to achieve higher output resistance, and consequently higher gain. However, this reduces the slew rate. Comparison between the two designs of [5] in Table 7, shows that if we halve the bias current, slew rate will be halved too, but gain increases only 16%. Therefore, if very high gain is desired, the proposed amplifier is superior, since it gives a better slew rate.

Among different amplifier topologies, single stage amplifiers have excellent frequency response [25]. The single stage amplifier in [5] has 20% less power consumption and drives a 10% larger load than the proposed amplifier, whereas they have a same settling time. However, the proposed amplifier has a higher settling accuracy due to its high DC gain, and similar to [5], does not become unstable when the load is increased. Shortly, the proposed amplifier has a higher gain and lower offset as compared to [5], whereas their power consumption, settling time and area are almost similar in the same process.

Unlike the two-stage amplifiers in [26, 27], the proposed two-stage amplifier does not become unstable if the load is increased, although it has two stages of gain. Moreover, compared to the amplifier in [27] which is implemented in 0.13um CMOS process, the proposed amplifier occupies 4 times less area, although implemented in 0.18um CMOS process.

Gain of the two-stage amplifier in [28] is 6 times lower than the proposed amplifier and [28] has lower slew rate. Regarding the later property, note that in [28], the load capacitance is 20 times smaller than in the proposed amplifier, whereas its slew rate is only 9.75 times larger. In other words, slew rate in [28] is less than half of this work. Moreover, it can become unstable if the load capacitance is increased.

High gain is especially important in applications such as high-accuracy sigma-delta modulators, or pipeline and flash analog to digital converters [11]. SC integrators used in the loop filter of sigma-delta modulators, need high open loop gain for accurate integration. Amplifiers used in these circuits, usually need 70-80dB gain to reduce phase error and other nonidealities [21, 29, 30].

Moreover, in mixed-mode circuits where the supply voltage experiences sever variations because of SC and other digital circuits, it is very important to have amplifiers with high PSRR [17, 31, 32].

In addition, accuracy of center frequency is a challenge for SC filters, because at high frequencies, they are usually sensitive to power supply, process, and bias current variations. Also, amplifiers in these circuits usually need high gain to obtain high Q [33].

Since the proposed amplifier has a high DC gain and also high $PSRR^+$ at high frequencies (near GBW), it is a good choice in the aforementioned applications. Moreover, its bias current being almost constant at different process corners, is another advantage for these applications.

Output voltage of sensors is usually in range of a few microvolts, and amplifiers with more than 40dB closed loop gain, 1% accuracy, cut off frequency of a few KHz, and low offset are usually required for boosting such signals to levels compatible with typical analog-to-digital converters [34]. The proposed amplifier is a good choice for such applications with 82.7dB gain, 22.3 KHz cut off frequency, and low offset.

## 5 Conclusions

In this paper, a novel structure for a two-stage high gain amplifier is presented. By applying a PCF around the second stage, effect of the pole at the output of the first stage is considerably reduced and the dominant pole happens at the output of the second stage. The proposed two-stage amplifier has been compared with other works, showing performance improvements in terms of DC gain and area. Moreover, in contrast to Miller compensated two-stage amplifiers whose $PSRR^+$ approaches 0dB around the GBW, in the proposed amplifier the $PSRR^+$ is more than 24dB at GBW and even improves if load capacitance is increased. Simple structure, low offset, good performance at presence of mismatch, high linearity, and also low sensitivity of stability to load capacitance are other advantages of the proposed amplifier.

## 6 References

1 Dadashi, A., Sadrafshari, Sh., Hadidi, Kh., Khoei, A.: 'Fast-settling CMOS Op-Amp with improved DC-gain' , *Analog Integr. Circuits Signal Process.*, 2012, 70, (3), pp. 283–292

2 Dadashi, A., Sadrafshari, Sh., Hadidi, Kh., Khoei, A.: 'An enhanced folded cascode Op-Amp using positive feedback and bulk amplification in 0.35 μm CMOS process' , *Analog Integr. Circuits Signal Process.*, 2011, 67, (2), pp. 213–222

3 Asloni, M., Hadidi, Kh., Khoei, A.: 'Design of a new folded cascode Op-Amp using positive feedback and bulk amplification' , *IEICE Trans. on Electronics*, 2007, E90-C, (6), pp. 1253–1257

4 Mottaghi-Kashtiban, M., Hadidi, Kh., Khoei, A.: 'Modified CMOS Op-Amp with improved gain and bandwidth' , *IEICE Trans. on Electronics*, 2006, E89-C, (6), pp. 775–780

5 S. Assaad, R., Silva-Martinez, J.: 'The Recycling Folded Cascode: A General Enhancement of the Folded Cascode Amplifier' , *IEEE J. Solid-State Circuits*, 2009, 44, (9), pp. 2535–2542

6 Li, Y.L., Han, K.F., Tan, X., Yan, N., Min, H.: 'Transconductance enhancement method for operational transconductance amplifiers' , *IET Electron. Lett.*, 2010, 46, (19), pp. 1321–1323

7 Yavari, M.: 'Single-stage class AB operational amplifier for SC circuits' , *IET Electron. Lett.*, 2010, 46, (14), pp. 977–979

8 Zhao, X., Fang, H., Xu, J.: 'A transconductance enhanced recycling structure for folded cascode amplifier', *Analog Integr. Circuits Signal Process.,* 2012, 72, (1), pp. 259-263

9 Razavi, B.: 'Design of analog CMOS integrated circuits', (McGraw-Hill, New York, USA, 2001)

10 Gray, P. R., Hurst, P. J., Lewis, S. H., Meyer, R. G.,: 'Analysis and Design of Analog Integrated Circuits', (Wiley, New York, USA, 2001, 4th edn.)

11 Baker, R.J.: 'CMOS circuit design, layout and simulation' (Wiley, 2008, 2nd edn.)

12 Aloisi, W., Palumbo, G., Pennisi, S., 'Design methodology of Miller frequency compensation with current buffer/amplifier', *IET Circuits, Devices & Systems*, 2008, 2, (2), pp.227-233

13 Rincon-Mora, G. A.: 'Active capacitor multiplier in Miller-compensated circuits', *IEEE J. Solid-State Circuits*, 2000, 35, (1), pp. 26–32 14 Hurst, P.J., Lewis, S.H., Keane, J.P., Aram, F., Dyer, K.C.: ‘Miller compensation using current buffers in fully differential CMOS two-stage operational amplifiers’, *IEEE Trans. Circuits Syst. I*, 2004, 51, (2), pp. 275-285

15 Yan, Z., Mak, P-I., Martins, R.P.: ‘Two Stage Operational Amplifiers: Power and Area Efficient Frequency Compensation for Driving a Wide Range of Capacitive Load’, *IEEE Circuits and Syst. Magazine*, 2011, 11, (1), pp. 26-42

16 Blakiewicz, G.: ‘Frequency compensation for two-stage operational amplifiers with improved power supply rejection ratio characteristic’, *IET Circuits, Devices & Systems*, 2010, 4, (5), pp. 458-467

17 Steyaert, M.S.J., Sansen, W.: ‘Power supply rejection ratio in operational transconductance amplifiers,’ *IEEE Trans. Circuits Syst. I*, 1990, 37, (9), pp.1077-1084

18 Kulej, T.: ‘0.5-V bulk-driven CMOS operational amplifier’, *IET Circuits, Devices & Systems*, 2013, 7, (6), pp. 352-360

19 Chatterjee, S., Tsividis, Y., Kinget, P.: ‘0.5-V analog circuit techniques and their application in OTA and filter design’, *IEEE J. Solid-State Circuits*, 2005, 40, (12), pp. 2373-2387

20 Vadipour, M.: ‘Capacitive feedback technique for wide-band amplifiers’, *IEEE J. Solid-State Circuits*, 1993, 28, (1), pp. 90-92

21 Lee, K-S., Kwon, S., Maloberti, F.: ‘A Power-Efficient Two-Channel Time-Interleaved ΣΔ Modulator for Broadband Applications’, *IEEE J. Solid-State Circuits*, 2007, 42, (6), pp. 1206-1215

22 R. kinget, P.: ‘Device Mismatch and Tradeoffs in the Design of Analog Circuits’, *IEEE J. Solid-State Circuits*, 2005, 40, (6), pp. 1212-1224

23 Pelgrom, M., H. Tuinhout, H., Vertregt, M.: ‘Transistor matching in analog CMOS applications’ , in *IEDM Tech. Dig.*, 1998, pp.915–918.

24 Yeh, Ta-H.,C.H Lin, J , Wong, Sh-Ch., Huang, H., Y.C Sun, J.: ‘Mis-match Characterization of 1.8V and 3.3V Devices in 0.18um Mixed Signal CMOS Technology’ *Proc. of the int. conference on microelectronic test structures(ICMTS)*, 2001, 14, pp. 77 –82.

25 Nang Leung, K., K. T. Mok, Ph.: ‘Analysis of Multistage Amplifier–Frequency Compensation’, *IEEE Trans. Circuits Syst. I*, 2001, 48, (9), pp.1041-1056

26 Taherzadeh-Sani, M., Hamoui, A.A.: ‘A 1-V Process-Insensitive Current-Scalable Two-Stage Opamp With Enhanced DC Gain and Settling Behavior in 65-nm Digital CMOS’, *IEEE J. Solid-State Circuits*, 2011, 46, (3), pp. 660-668

27 Figueiredo, M., Santos-Tavares, R., Santin, E., Ferreira, J., Evans, G., Goes, J.: ‘A Two-Stage Fully Differential Inverter-Based Self-Biased CMOS Amplifier With High Efficiency’, *IEEE Trans. Circuits Syst. I*, 2011, 58, (7), pp.1591-1603

28 Perez, A.P., Maloberti, F.: ‘Performance enhanced op-amp for 65nm CMOS technologies and below’, IEEE Int. Symp. on Circuits and Systems (ISCAS), Seoul, South Korea, 2012, pp. 201-204

29 Yao, L., S. J. Steyaert, M., Sansen, W.: ‘A 1-V 140-uW 88-dB Audio Sigma-Delta Modulator in 90-nm CMOS’, *IEEE J. Solid-State Circuits*, 2004, 39, (11), pp. 1809-1818 30 Geerts, Y., Marques, A. M., S. J. Steyaert, M., Sansen, W.: 'A 3.3-V, 15-bit, Delta–Sigma ADC with a Signal Bandwidth of 1.1 MHz for ADSL Applications', *IEEE J. Solid-State Circuits*, 1999, 34, (7), pp. 927-936

31 Loikkanen, M., Keranen, P., Kostamovaara, J.: 'Single supply high PSRR class AB amplifier', *IET Electron. Lett.*, 2008, 44, (2), pp. 70–71

32 Sobhy, E.A., Hoyos, S., Sanchez-Sinencio, E.: 'High-PSRR low-power single supply OTA', *IET Electron. Lett.*, 2010, 46, (5), pp. 337–338

33 Nicollini, G., Moretti, F., Conti, M.: 'High-Frequency Fully Differential Filter Using Operational Amplifiers Without Common-Mode Feedback', *IEEE J. Solid-State Circuits*, 1989, 24, (3), pp. 803-813

34 F. Witte, J., A. A. Makinwa, K., H. Huijsing, J.: 'A CMOS Chopper Offset-Stabilized Opamp', *IEEE J. Solid-State Circuits*, 2007, 42, (7), pp. 1529-1535